\documentclass[%
 reprint,
 amsmath,amssymb,
 aps,
]{revtex4-2}

\usepackage{graphicx}
\usepackage{dcolumn}
\usepackage{bm}

\usepackage{caption}
\usepackage{subcaption}
\usepackage{amssymb}
\usepackage{braket}
\usepackage{dsfont}

\begin{document}

\preprint{APS/123-QED}

\title{Rectification induced by geometry in two-dimensional quantum spin lattices}

\author{Alessandra Chioquetta}
\email{achioquetta@fis.dout.ufmg.br}
\affiliation{Departamento de Física–Instituto de Ciências Exatas, Universidade Federal de Minas Gerais, CP 702,
30.161-970 Belo Horizonte MG, Brazil}
\author{Emmanuel Pereira}%
\affiliation{Departamento de Física–Instituto de Ciências Exatas, Universidade Federal de Minas Gerais,
CP 702, 30.161-970 Belo Horizonte MG, Brazil}

\author{Gabriel T. Landi}
\affiliation{Instituto de Física da Universidade de São Paulo, 05314-970 São Paulo, Brazil}

\author{Raphael C. Drumond}
\affiliation{Departamento de Matemática, Universidade Federal de Minas Gerais, Belo Horizonte, MG, Brazil}



\date{\today}

\begin{abstract}
We address the role of geometrical asymmetry in the occurrence of spin rectification in two- dimensional quantum spin chains subject to two reservoirs at the boundaries, modeled by quantum master equations. We discuss the differences in the rectification for some one- dimensional cases, and present numerical results of the rectification coefficient R for different values of the anisotropy parameter of the XXZ model, and different configurations of boundary drives, including both local and non-local dissipators. Our results also show that geometrical asymmetry, along with inhomogeneous magnetic fields, can induce spin current rectification even in the XX model, indicating that the phenomenon of rectification due to geometry may be of general occurrence in quantum spin systems.
\end{abstract}

\maketitle


\section{Introduction}

If we couple two reservoirs to a system inducing a flux through it (for instance, of heat or particles), and if the total system has the property of being relaxing, it will reach a non-equilibrium steady state (NESS), characterized by a constant flow. Then, if we reverse the reservoirs and the flow induced by the exchanged reservoirs has a different magnitude of the first configuration, we say that the system exhibits rectification.

Besides the familiar rectification of electric currents, thermal rectification was observed as well for the first time by Starr (1936) in copper oxide rectifiers \cite{brattain1951copper,starr1936copper} and has gained more and more attention nowadays, experimentally and theoretically, largely due to the possibility of experimentally realizing  thermal circuits and diodes \cite{li2004thermal,nefzaoui2014radiative,hu2009thermal,yan2009control,avila2013thermal,terraneo2002controlling,li2012colloquium, kobayashi2009oxide}. The phenomenon can also be observed in quantum many-body systems for energy and magnetization currents~\cite{schuab2016energy,landi2014flux,balachandran2018perfect}. Understanding the transport properties of those systems can lead to technological advances and, furthermore, bring new insights for fundamental physics.

Even though rectification is a well-known phenomenon, the essential elements for its occurrence are not yet fully understood. Undoubtedly, an asymmetric component is required, but not any type of asymmetry is sufficient \cite{pereira2011sufficient}. 
In phononics, for example, a considerable effort has been dedicated to the proposal of an efficient and feasible thermal diode. Systems given by the sequential coupling of two or three segmented parts \cite{terraneo2002controlling}, graded systems \cite{pereira2010graded}, i.e., devices in which the structure changes gradually in space, arrangements involving long range interactions \cite{chen2015ingredients} and other mechanisms have been recurrently investigated. A graded thermal diode has been already experimentally built \cite{chang2006solid}, given by a carbon and boron nitride nanotube, externally and asymmetrically coated with heavy molecules. Unfortunately, its thermal rectification factor is very small. Geometrical arrangements, in particular for some intricate graphene models, have been also considered as a possible mechanism for thermal rectification. For example, a graphene nanoribbon with a two-dimensional trapezoidal shape is studied in Ref.\cite{wang2014phonon}; graphene Y junctions are considered in Ref.\cite{zhang2011thermal}; and graphene nanoribbons in triangular shapes are investigated in Ref.\cite{yang2009thermal}.

A subject of currently increasing attention is the study of the transport laws at quantum scale. Motivated by the advances of quantum thermodynamics and the possibility of building quantum devices due to the progress of nanotechnology, several works are devoted to the theme. In particular, quantum spin chains described by Heisenberg and XXZ models are exhaustively visited. These systems are related to problems in different areas: cold atoms, quantum information, condensed matter, optics, etc. 

Important results about rectification in one-dimensional quantum spin chains have being obtained \cite{balachandran2018perfect, landi2014flux,zhang2009reversal,oliveira2020one,van2011rectification}. For example, in Ref.\cite{schuab2016energy} and \cite{pereira2017rectification} it was shown that for the XXZ model, a graded interaction induces rectification of energy currents, with no need of a graded magnetic field. Conversely, for magnetic currents, as discussed in Ref.\cite{landi2014flux}, the XXZ model under a graded magnetic field is enough for the existence of rectification. However, for the XX model this configuration does not suffice.

When extended to two-dimensional lattices, not many studies on rectification in spin systems have been reported, so 
in this work we focus on the theme and consider four distinct geometries for open quantum spin lattices governed by the XXZ model. In  order to explore the role of the geometries on spin transport, two cases with geometrical asymmetry and two symmetrical cases are examined, each one with different number of coupled reservoirs. The dynamics are described by the Lindblad master equation \cite{breuer2002theory,prosen2011open,gorini1976completely} and the solution for the steady state is obtained through vectorization of the time evolution operator. We analyse the rectification behavior for different values of the anisotropy parameter $\Delta$ and three configurations of an external magnetic field: one homogeneous and two non-homogeneous changing along the lattice. We show  numerical results for the occurrence of rectification in symmetrical XXZ and asymmetrical XXZ and XX ($\Delta=0$) models under non-homogeneous magnetic fields.

This work is organized as follows: In Sec. \ref{sec2} we begin describing the XXZ model, the symmetric and asymmetric geometries that we consider and also the master equation approach for the dynamics of the open systems. In Sec. \ref{sec3} we discuss how to evaluate spin currents and quantify rectification in two-dimensional XXZ spin lattices. Also, we present the solution for the steady state of Lindblad master equation through vectorization and time evolution operator. Finally, in Sec. \ref{sec4} we show numerical results for all geometries and magnetic fields considered. This section contains the most important results of our work, particularly for the asymmetric XX model, which presents rectification under a non-homogeneous magnetic field.


\section{Model} \label{sec2}
\begin{figure}[h]
\begin{center}
   \begin{subfigure}{.25\textwidth}
  \includegraphics[width=1.\linewidth]{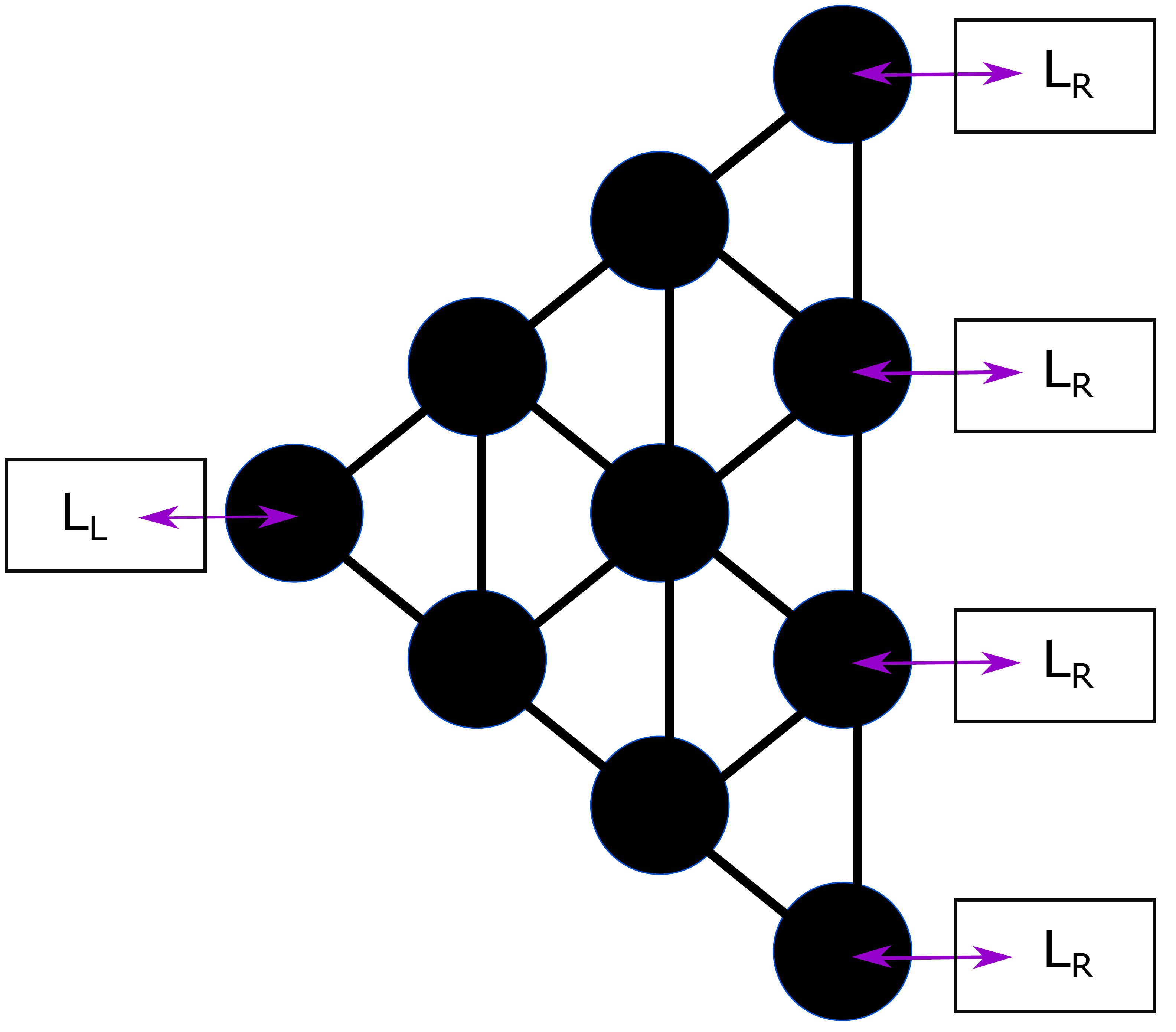} 
  \caption{}
    \label{g1}
    \end{subfigure}
    \qquad \begin{subfigure}{.25\textwidth}
  \includegraphics[width=1.\linewidth]{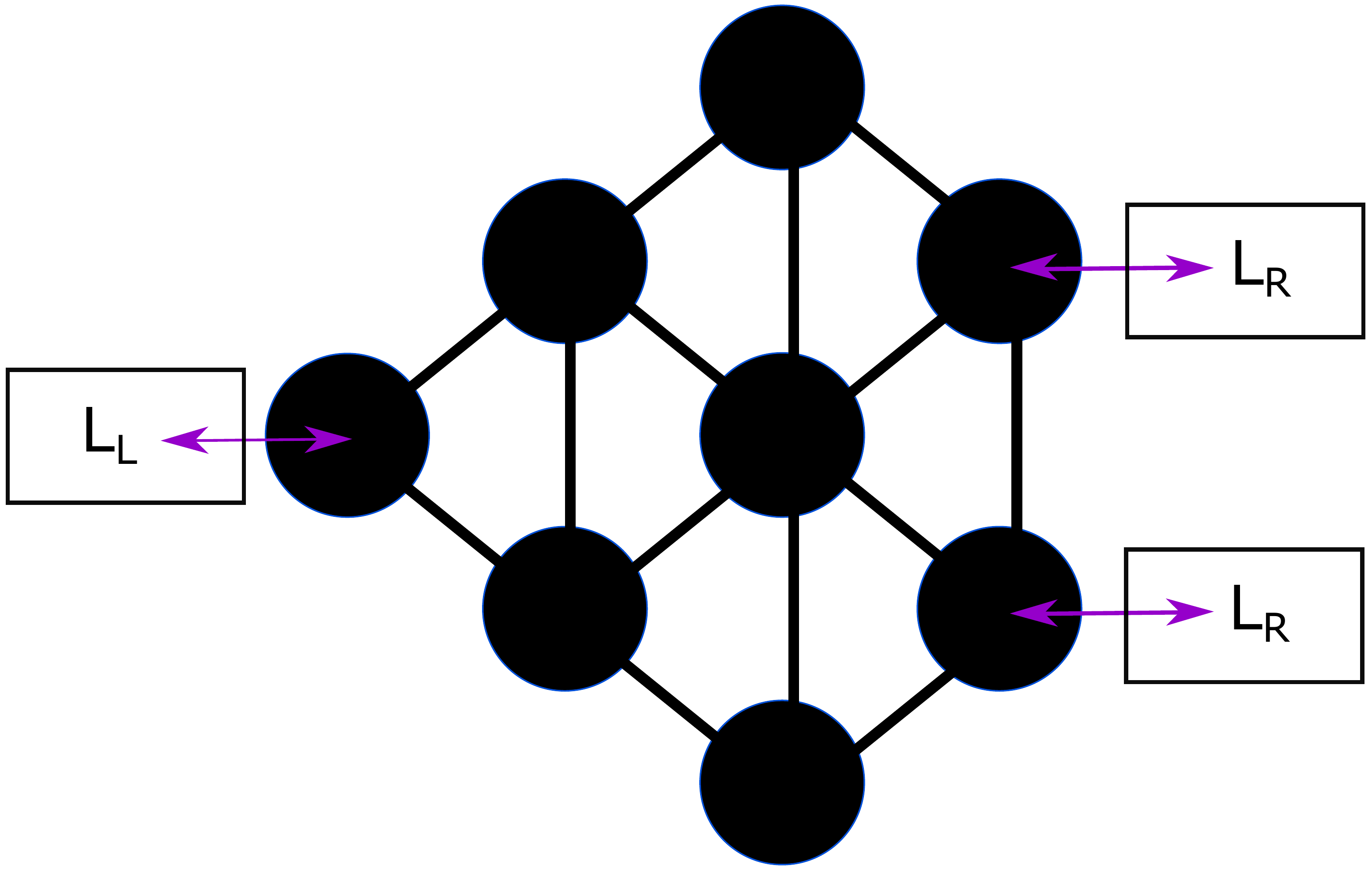} 
  \caption{ }
    \label{g2}
    \end{subfigure}
    \qquad \begin{subfigure}{.25\textwidth}
  \includegraphics[width=1.\linewidth]{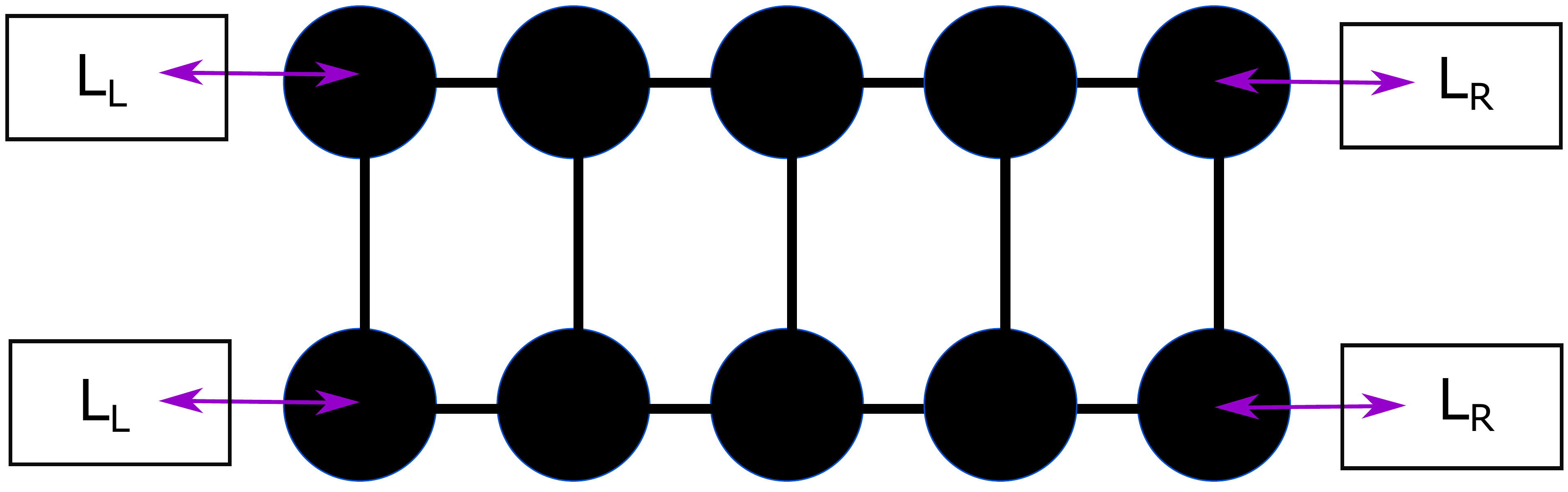} 
  \caption{ }
    \label{g3}
    \end{subfigure}
    \qquad \begin{subfigure}{.25\textwidth}
  \includegraphics[width=1.\linewidth]{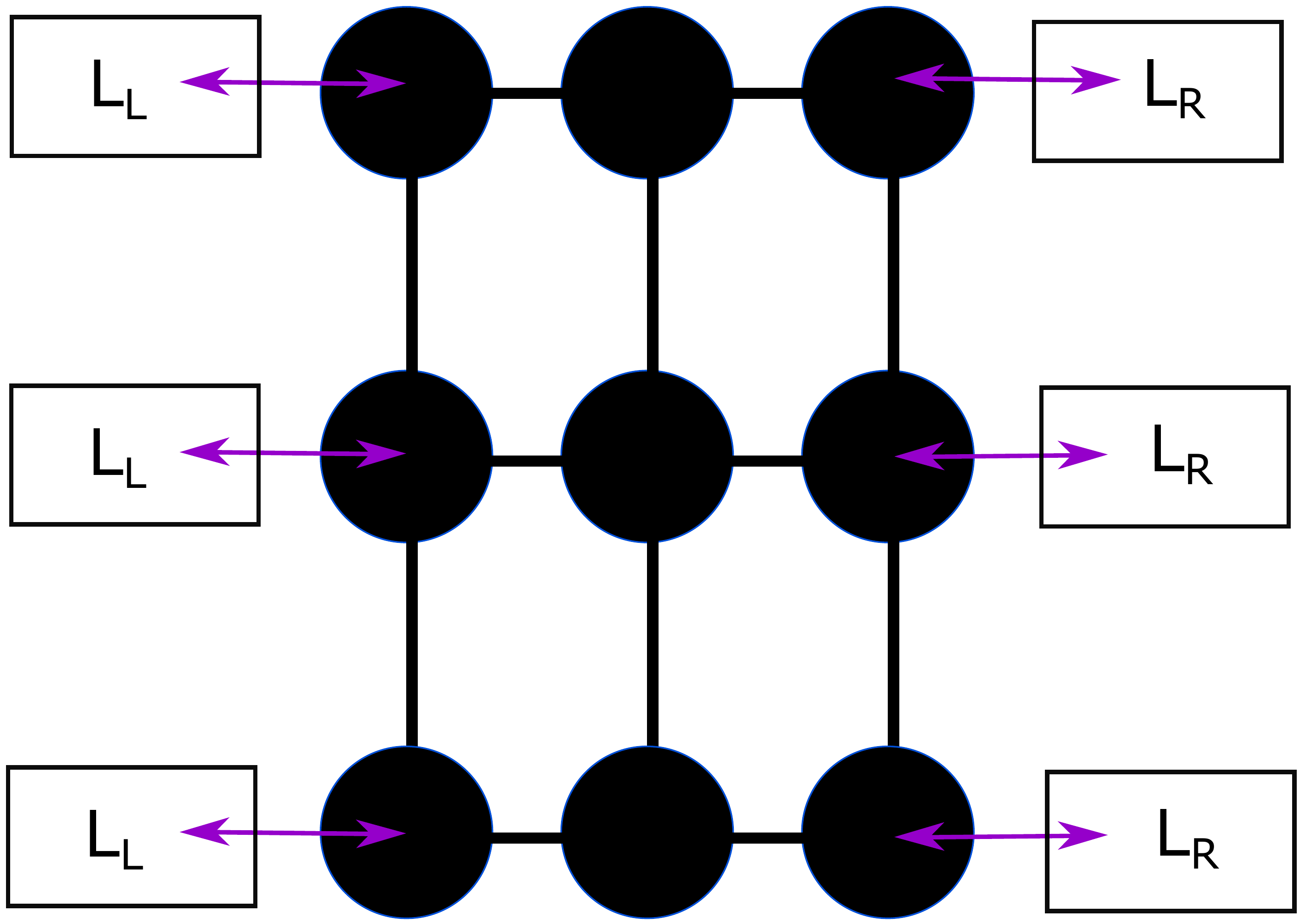} 
  \caption{ }
    \label{g4}
    \end{subfigure}
    \end{center}
    \caption{ Four lattices considered: Two with geometrical asymmetry and two symmetric. In (a) asymmetric geometry with ten sites, one reservoir on the left side and four on the right side; (b) asymmetric geometry with eight sites, one reservoir on the left and two on the right side; (c) symmetric geometry with ten sites, two reservoir on the left and two on the right side; (d) symmetric geometry with nine sites, three reservoirs on the left and three on the right side.}
     \label{geometries}
\end{figure}

We consider four different geometries of two-dimensional spin-$\frac{1}{2}$ lattices coupled to magnetic reservoirs: two with asymmetric geometry and two symmetric, as represented in the Fig. \ref{geometries}. The Hamiltonian that describes the lattices was chosen to be in the XXZ form:
\begin{equation}
     H = \sum_{\langle i,j \rangle}[ \alpha( \sigma_{i}^{x} \sigma_{j}^{x} +  \sigma_{i}^{y} \sigma_{j}^{y}) +\Delta \sigma_{i}^{z} \sigma_{j}^{z}) ] + \sum_{i}^N h_{i} \sigma_{i}^{z},
\end{equation}
where $\sigma_{k}^{a}$ with $a \in x,y,z$ are the Pauli matrices, the sum $\langle i,j \rangle$ refers to sites that interact with each other,  and $h_{i}$ is the magnetic field acting on site $i$. In this work we set $\hbar =1$ and fix $\alpha = 1$.


For the dynamics we consider a local Lindblad equation approach, where the time evolution of the density matrix $\rho$ is given by the Lindblad master equation (LME):

\begin{equation}
    \mathcal{L}(\rho) = \frac{d}{dt} \rho = -i[H,\rho] + D_{L}(\rho) + D_{R}(\rho).
    \label{lindblad}
\end{equation}
The dissipative parts $D_{L}$ and $D_{R}$ are given by:

\begin{equation}
   D_{\beta}(\rho) = \sum_{j \in \beta} \sum_{s=\pm} \gamma_j (L_{j,s} \rho  L_{j,s}^{\dagger} - \frac{1}{2} \{L_{j,s}^{\dagger} L_{j,s}, \rho\} ),
\end{equation}
where $\{ \cdot\}$ is the anticommutator, $\beta = L,R$ are the sites coupled to the reservoirs on the left and right side respectively (see Fig. \ref{geometries}), $\gamma_{j}$ is a positive constant that will be specified for each geometry based on the number of coupled reservoirs and
\begin{equation}
    L_{j,\pm} = \sqrt{1 \pm f_{j}} \;\sigma_{j}^{\pm}
\end{equation}
describes the coupling between sites and reservoirs, where $\sigma_{j}^{\pm} = \frac{\sigma_{j}^{x} \pm i \sigma_{j}^{y}}{2}$ are the spin creation and annihilation operators acting on site $j$.   The parameter $f_{j}$ can be interpreted as the expectation value of the magnetization of an extra spin that is not part of the lattice coupled to the site $j$. When $f_{j \in L} \neq f_{j \in R}$ the system will evolve to a non-equilibrium steady state (NESS), which is characterized by a constant flow of magnetization through the lattice. In this work we set:
\begin{equation}
    f_{j \in L} = -f_{j \in R} = f,
\end{equation}
and $f=1$, which represents reservoirs with fixed magnetization as only up or down, not a mixture of them.


\section{Spin currents and Rectification} \label{sec3}

The spin currents may be found through evaluating the time variation of the local magnetization $\langle \sigma_{k}^{z}\rangle$. For sites that are not coupled to reservoirs it can be shown that:
\begin{equation}
     \frac{d}{dt} \langle \sigma_{k}^{z} \rangle =  \sum_{j} J_{kj}
     \label{corrente}
 \end{equation}
where the sum in $j$ are for sites that interact with site $k$ and
\begin{equation}
    J_{kj} = 2 \alpha \langle  \sigma_{k}^{x} \sigma_{j}^{y} -\sigma_{k}^{y} \sigma_{j}^{x} \rangle
\end{equation}
is the current from site $k$ to $j$. This form for the current can be obtained through the LME and continuity equation.

\begin{figure}
    \centering
    \includegraphics[scale=0.2]{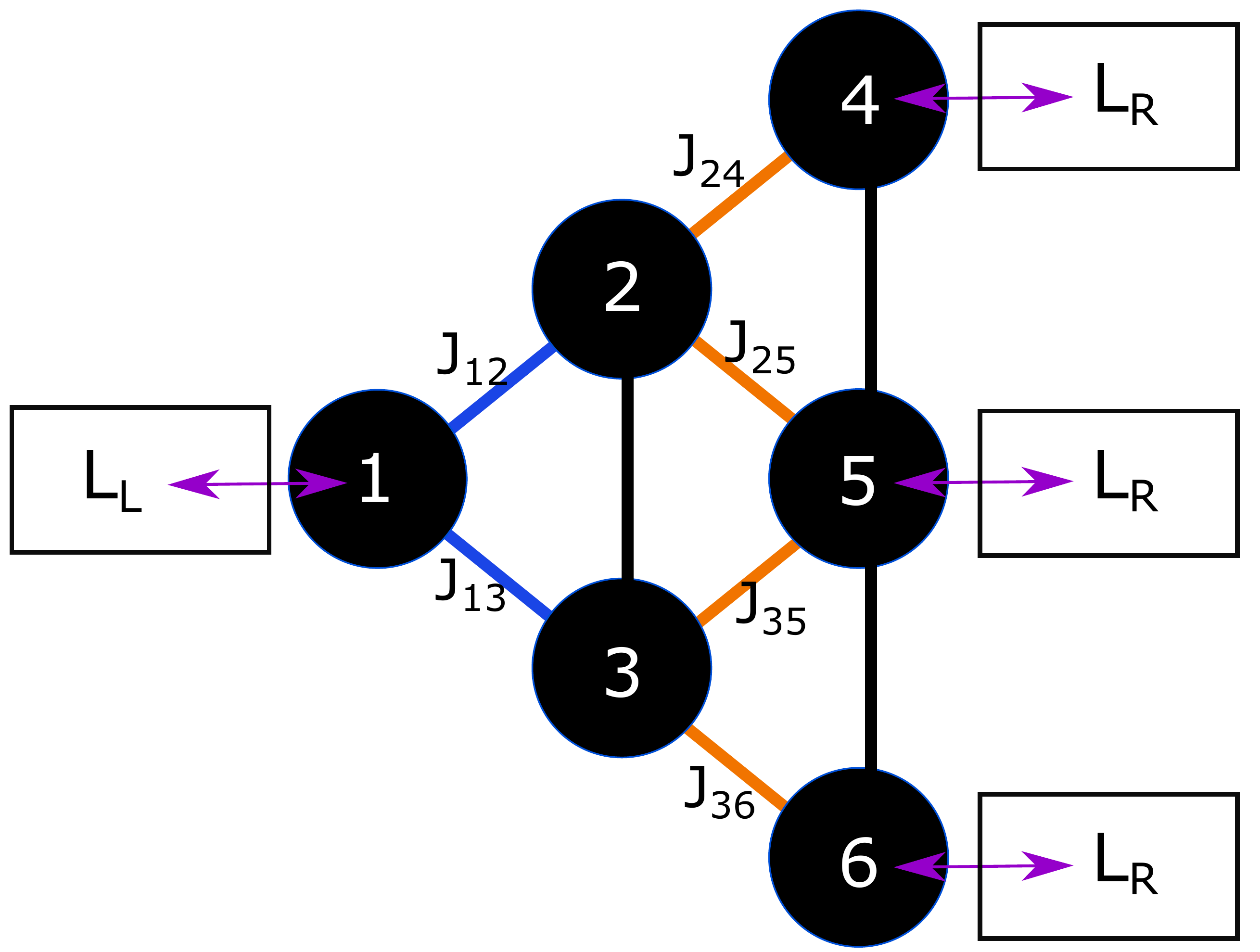}
    \caption{In the steady state the sum of the currents in blue must be the equal to the sum of the currents in orange. The vertical currents cancel themselves when the system reaches NESS.}
    \label{ness}
\end{figure}

When the system reaches the NESS we have $\frac{d}{dt}\langle \sigma_{k}^{z} \rangle = 0$ for every $k$. For one-dimensional cases we would obtain that the currents must be the same between each site, but for the two-dimensional cases we can consider the sum in a column of sites. If we have a triangular lattice with 6 sites, as in Fig. \ref{ness} and making the sum on sites 2 and 3, for example, we can obtain the relation:
\begin{equation}
J_{12}+J_{13} = J_{24}+J_{25}+J_{35}+J_{36} \equiv J, 
 \label{sumsteady}
\end{equation}
so the sum of the currents $J$ is homogeneous through the lattice. For different geometries the relation is analogous: the sum of the currents on the left side of a column of sites must be the same as the sum on the right side. This can be interpreted as the sum of the currents that ``goes in'' the column must be the same that ``goes out''. In this work the sum $J$ is considered to verify and quantify how much rectification the systems present.


To measure how much rectification a system has, we must compare $J(f)$ with $J(-f)$, where $J(f)$ is the steady state current and $J(-f)$ is the steady state current when the reservoirs are reversed, represented by the sign change $f \rightarrow -f$. Thus, we can define a rectification coefficient as \cite{landi2014flux}:
\begin{equation}
    R = \frac{J(f)+J(-f)}{J(f)-J(-f)}.
\end{equation}
If $R=0$ there is no rectification and if $R= \pm 1$ the system behaves as a perfect insulator in one direction, characterizing total rectification. The sign of $R$ only indicates in which direction the current is more intense. 

To evaluate the currents $J(f)$ and $J(-f)$ we must find the NESS solution first, where $\mathcal{L}(\rho) = 0$ is satisfied. For this calculation we start employing the vectorization method, which consists in converting a matrix in a column vector. For $2 \times 2$ matrices, the operation $\text{vec}(\cdot)$ is given by \cite{landi2014flux}: 
\begin{equation}
    \text{vec} \begin{pmatrix} a \;\; b \\ c \;\; d \end{pmatrix} = \begin{pmatrix} a \\ c \\ b \\ d \end{pmatrix} ,
\end{equation}
and for any matrices A, B and C, the identity
\begin{equation}
     \text{vec} (ABC) = (C^{\top}\otimes A) \text{vec} (B)
\end{equation}
may be verified.

We define the vectorization of the density matrix as
\begin{equation}
   \text{vec} (\rho) \equiv  \ket{\rho}, 
\end{equation}
and then, the vectorized terms of the Lindblad master equation can be written as:

\begin{equation}
    \begin{split}
        \text{vec}(-i[H,\rho]) &= -i(\mathds{1} \otimes H - H^{\top} \otimes \mathds{1})\ket{\rho}, \\
         \text{vec}(L_{j} \rho L_{j}^{\dagger}) &= (L_{j}^{*} \otimes L_{j}) \ket{\rho}, \\
        \text{vec}(L_{j}^{\dagger} L_{j} \rho) &= (\mathds{1}\otimes L_{j}^{\dagger} L_{j}) \ket{\rho},\\
        \text{vec}(\rho L_{j}^{\dagger} L_{j}) &= ((L_{j}^{\dagger} L_{j})^{\top} \otimes \mathds{1}) \ket{\rho}.
    \end{split}
\end{equation}

Thus, the LME can be written as
\begin{equation}
    \frac{d}{dt}\ket{\rho} = W \ket{\rho},
\end{equation}
and the solution for $\ket{\rho (t)}$ is given by time evolution operator:
\begin{equation}
    \ket{\rho(t)} = e^{Wt} \ket{\rho(0)}.
\end{equation}

Since the system under consideration has the relaxing property, the steady-state is reached when $t \rightarrow \infty$:
\begin{equation}
    \frac{d}{dt}\ket{\rho(t \to \infty)} = \frac{d}{dt}\ket{\rho_{ss}} = 0,
    \label{steadystate}
\end{equation}
hence we can write the NESS solution as:
\begin{equation}
    \ket{\rho_{ss}} = \lim_{t \to \infty} e^{Wt} \ket{\rho (0)}.
\end{equation}

The vectorized density matrix $\ket{\rho}$ has length $2^{2N}$ and, therefore, the time evolution approach is used instead of direct diagonalization, due to the large amount of sites on the lattices being considered and computational power restriction. 

We estimate $t=10^{3}$ (in $1/\alpha$ units) as enough time to the systems under consideration reach a state that is essentially in the NESS, and to ensure this the property (\ref{sumsteady}) is verified in every case, adapting it to each geometry.


\section{Results} \label{sec4}

For the four geometries considered under a homogeneous magnetic field (same magnetic field applied in each site) $h=1.0$, as well as the absence of it, rectification is not observed. But when we consider a non-homogeneous magnetic field, differences between asymmetric and symmetric geometries show up.

\begin{figure}[h]
   \begin{subfigure}{.25\textwidth}
  \includegraphics[width=1.1\linewidth]{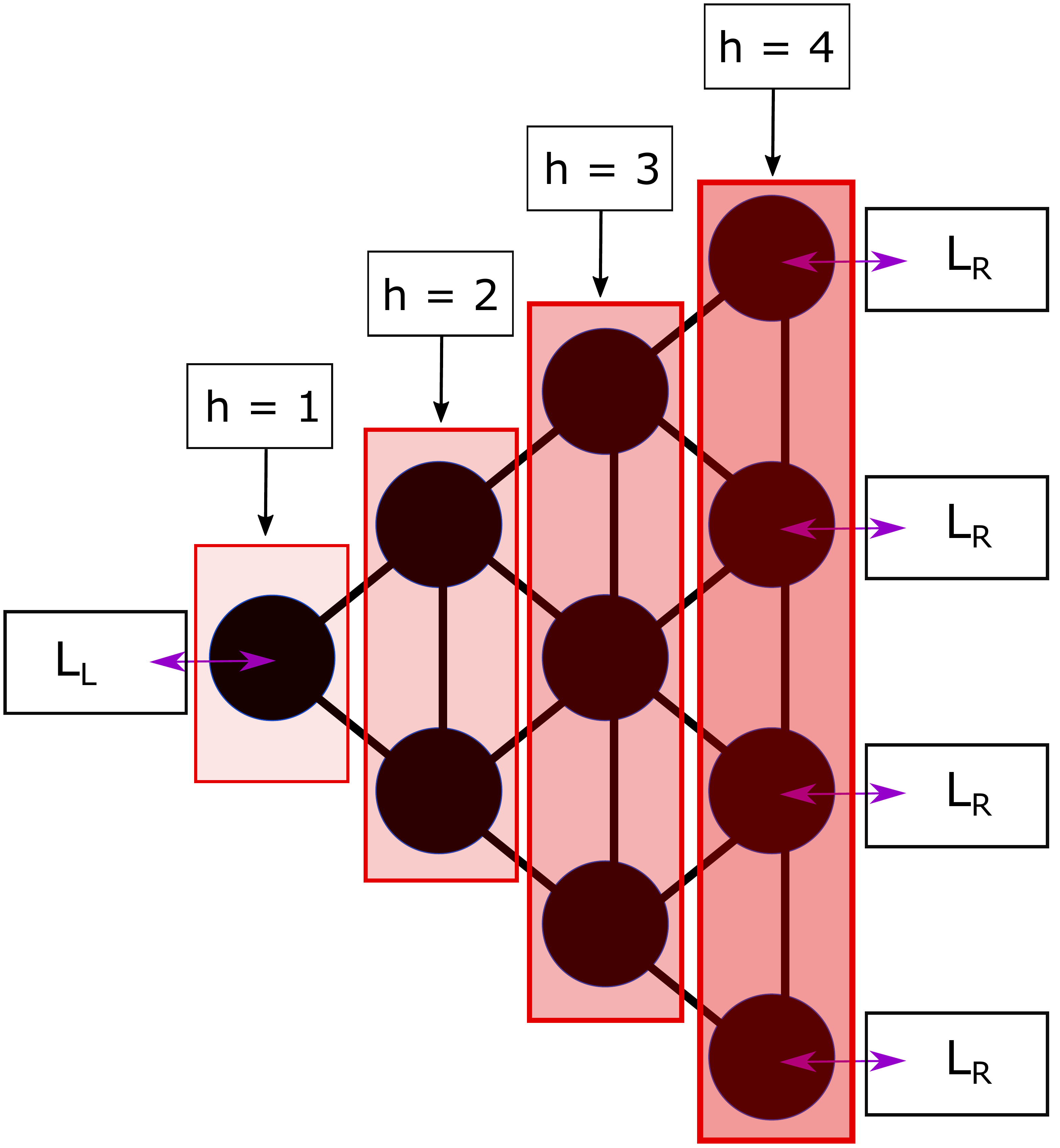} 
  \caption{}
    \label{g1c1}
    \end{subfigure}
    \qquad \begin{subfigure}{.25\textwidth}
  \includegraphics[width=1.1\linewidth]{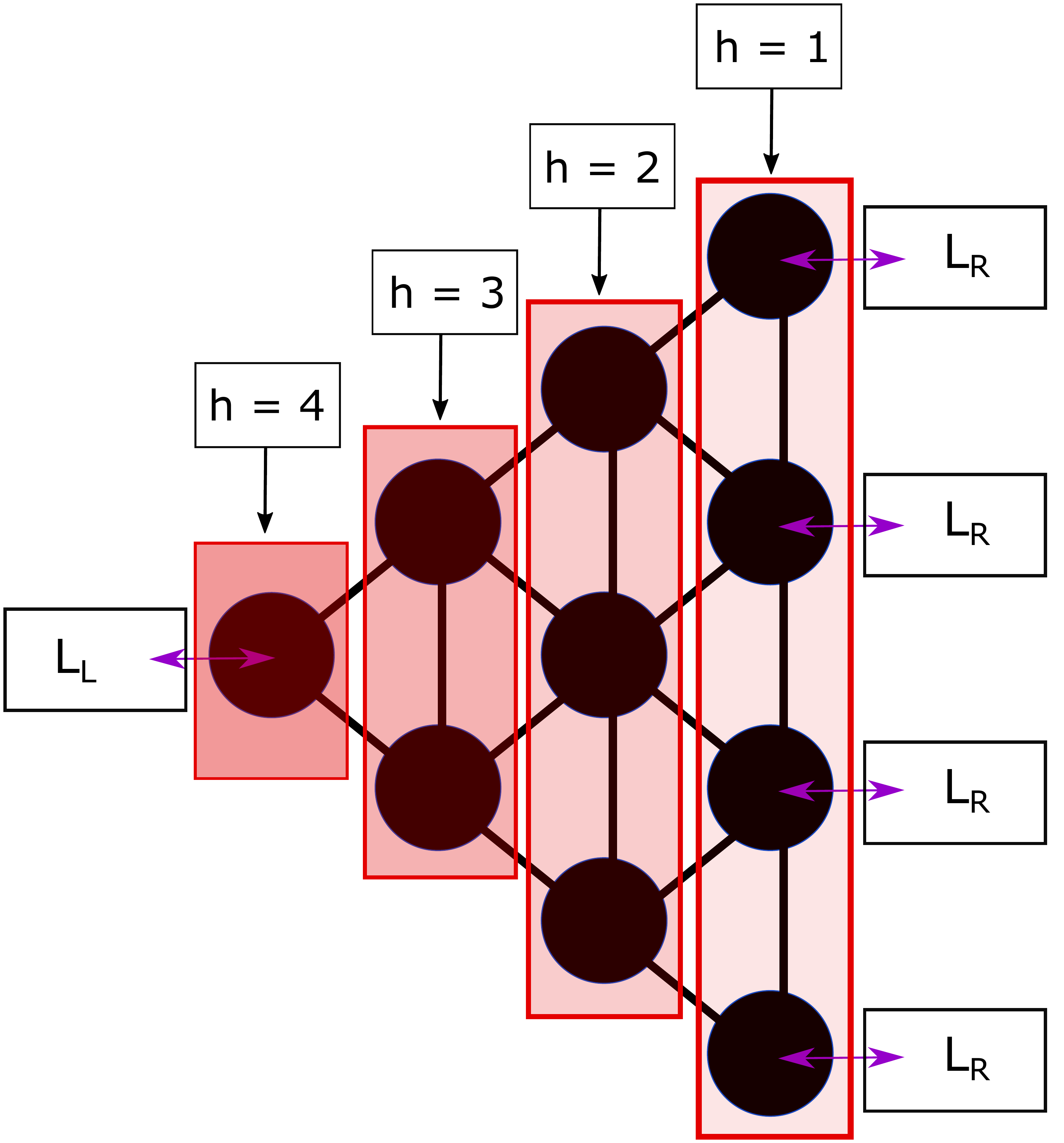} 
  \caption{}
    \label{g1c2}
    \end{subfigure}
    \caption{Field configurations 1 and 2 for the triangular lattice with 10 sites. (a) Field configuration 1 of the triangular lattice with 10 sites: magnetic field varies from $h=1$ to $h=4$ from left to right; (b) field configuration 2 of the triangular lattice with 10 sites: magnetic field varies from $h=1$ to $h=4$ from right to left}
     \label{fields}
\end{figure}

Two configurations of non-homogeneous magnetic field are investigated: The first one (Field Configuration $1$) $h$ increases in one unit for each column of sites from left to right, and the second case (Field Configuration $2$) $h$ increases  in one unit from right to left. For the asymmetrical lattice with $10$ sites the field configurations are represented in the Fig. \ref{fields} and the other lattices follow the same idea, depending only on the number of columns they have.


\begin{figure}[h]
    \centering
    \includegraphics[scale=0.26]{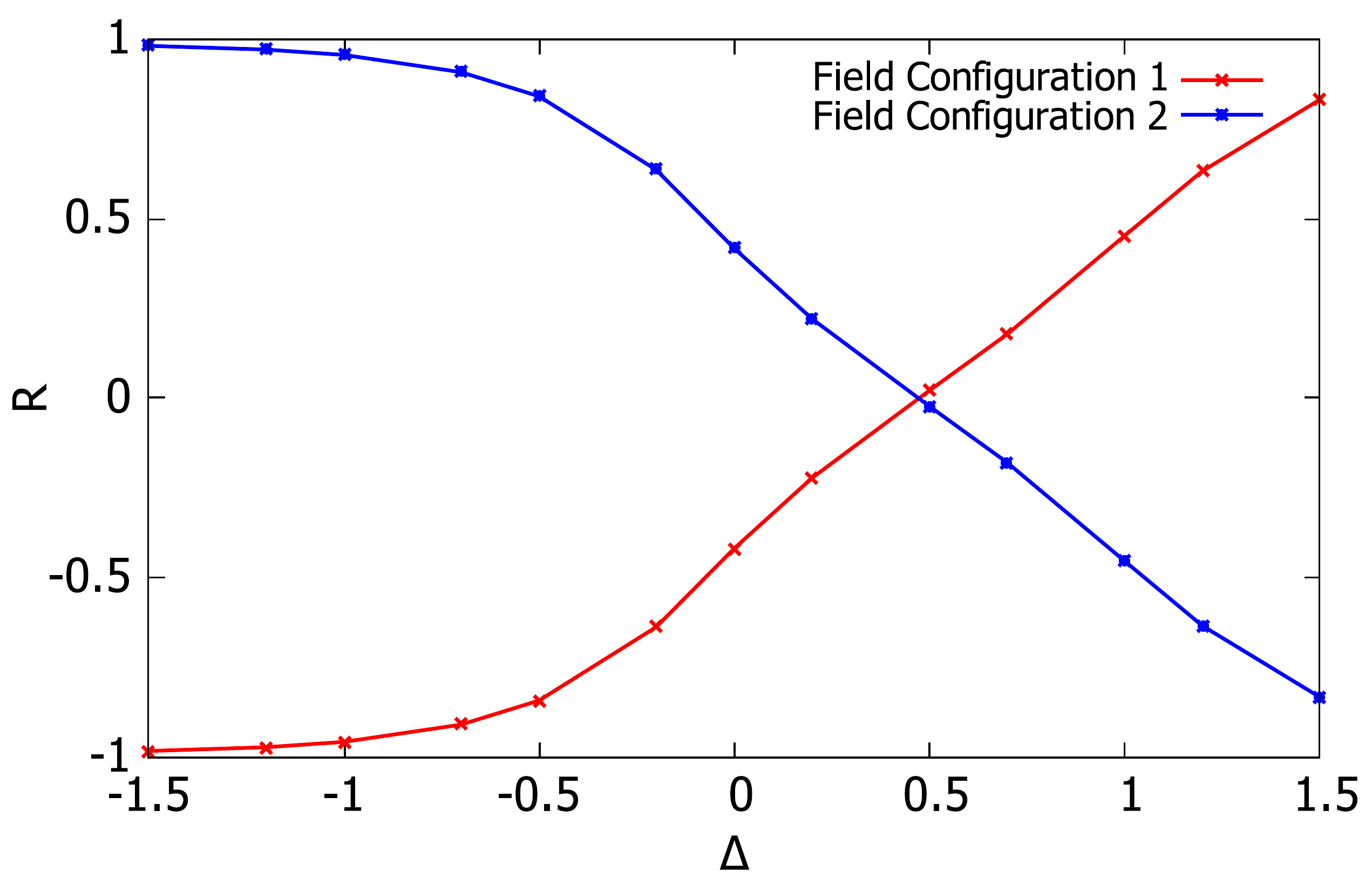}
    \caption{Rectification coefficient $R$ vs $\Delta$ for the asymmetrical lattice with ten sites (Fig. \ref{g1}), under two configurations of non-homogeneous magnetic fields (Fig. \ref{fields}). }
    \label{g1plot}
\end{figure}

\begin{figure}[h]
    \centering
    \includegraphics[scale=0.26]{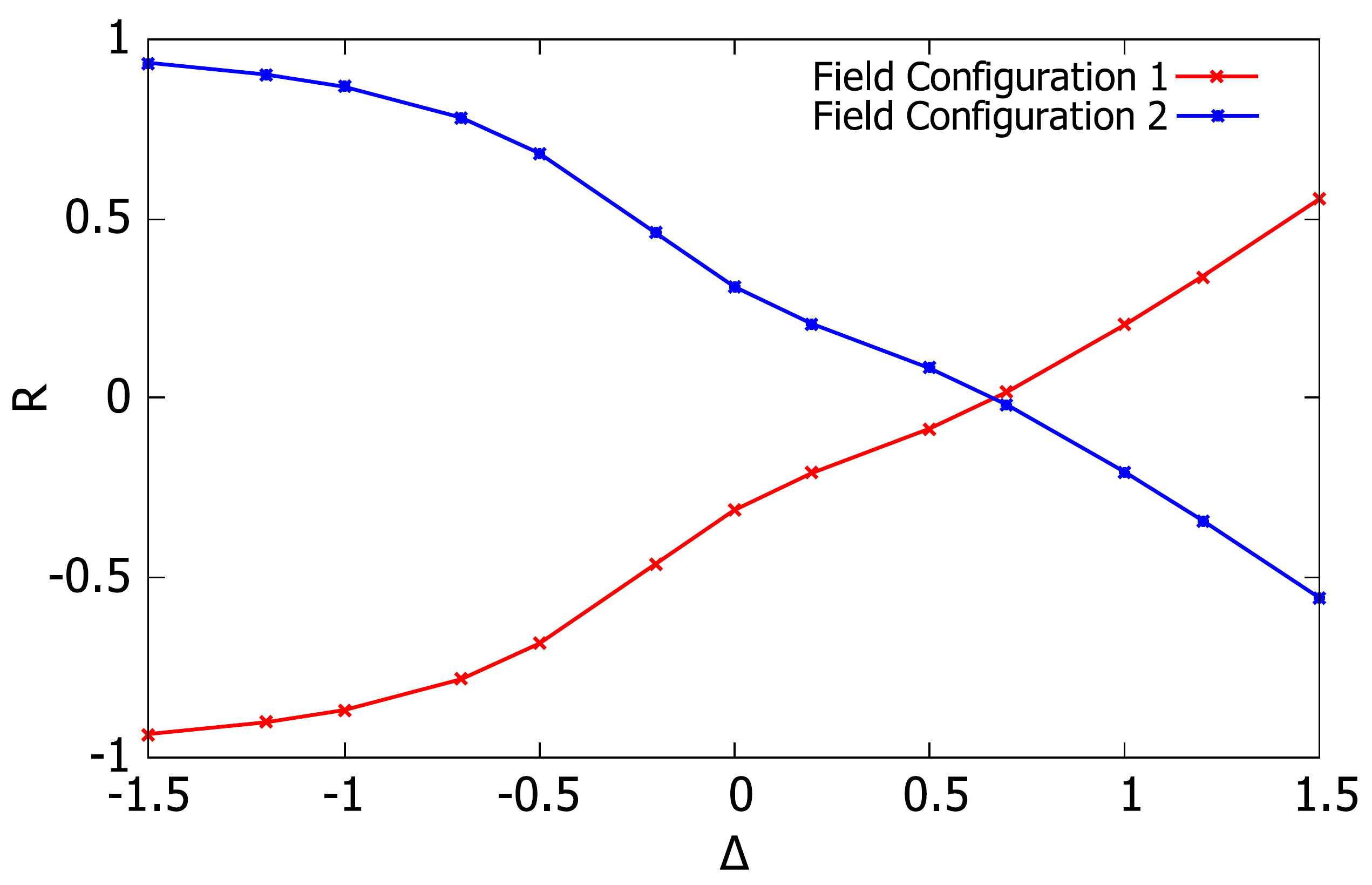}
    \caption{Rectification coefficient $R$ vs $\Delta$ for the asymmetrical lattice with eight sites (Fig. \ref{g2}), under two configurations of non-homogeneous magnetic fields. }
    \label{g2plot}
\end{figure}

The results for the first lattice (Fig. \ref{g1}) with geometrical asymmetry, under a magnetic field varying from $h=1.0$ to $h=4.0$ and coupling parameter $\gamma_{j \in R} = \gamma_{j \in L}/4 = 1.0/4$, are in  Fig. \ref{g1plot}. For the second lattice (Fig. \ref{g2}), under a a field varying from $h=1.0$ to $h=4.0$ as well and $\gamma_{j \in R} = \gamma_{j \in L}/2=1.0/2$, are in Fig. \ref{g2plot}.
Differently from one-dimensional chains, the XX model ($\Delta =0$) shows rectification in the presence of a non-homogeneous magnetic field and have a point where $R=0$ for $\Delta > 0$ in both geometries, that seems to be a symmetry point. Another interesting detail is the change only on the sign of the rectification coefficient $R$ when the magnetic field configuration changes, which is unexpected since the total magnetic field applied is not the same.  

\begin{figure}[h]
    \centering
    \includegraphics[scale=0.26]{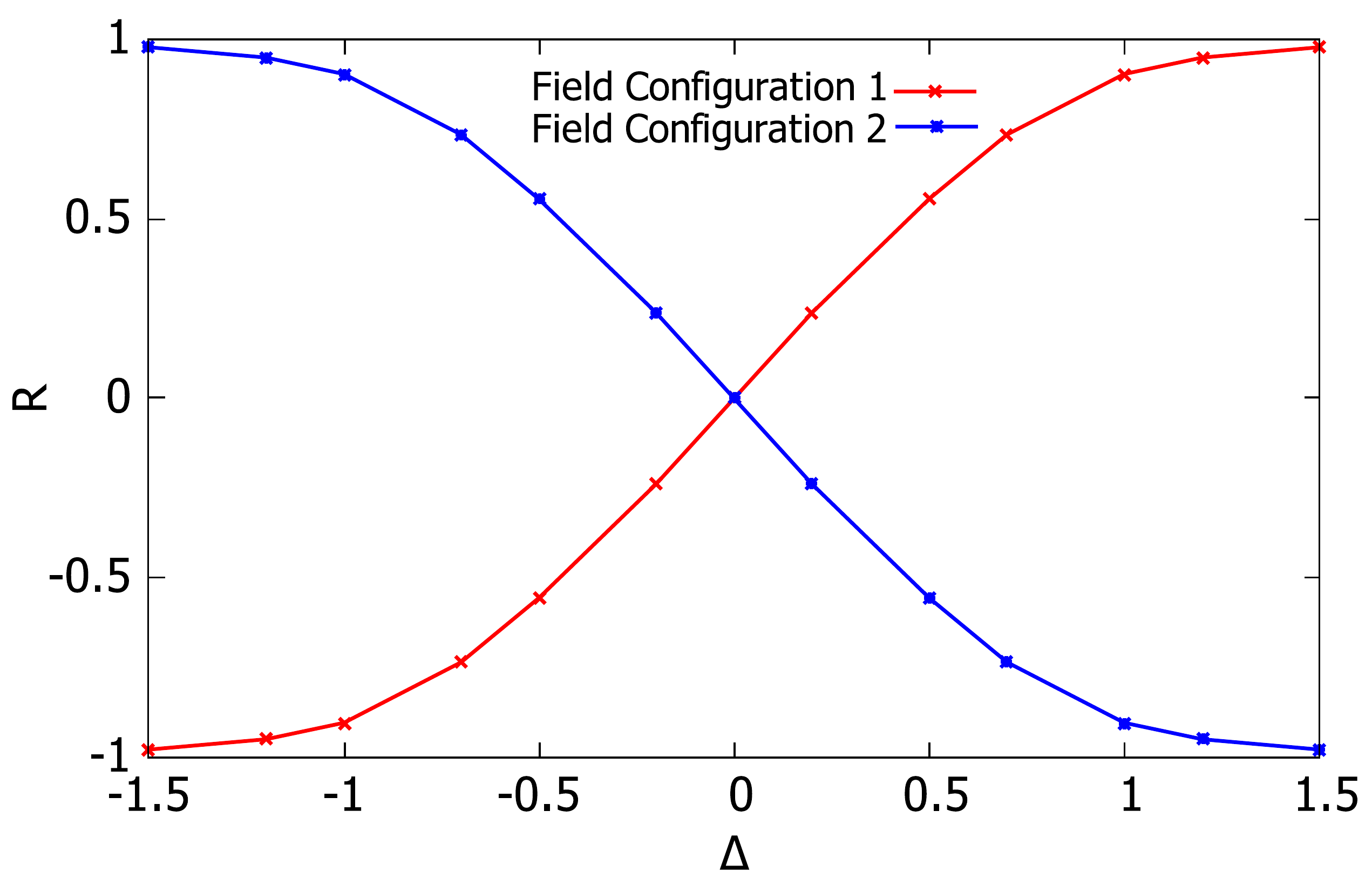}
    \caption{Rectification coefficient $R$ vs $\Delta$ for the symmetrical lattice with ten sites (Fig. \ref{g3}), under two configurations of non-homogeneous magnetic fields.}
    \label{g3plot}
\end{figure}

\begin{figure}[h]
    \centering
    \includegraphics[scale=0.26]{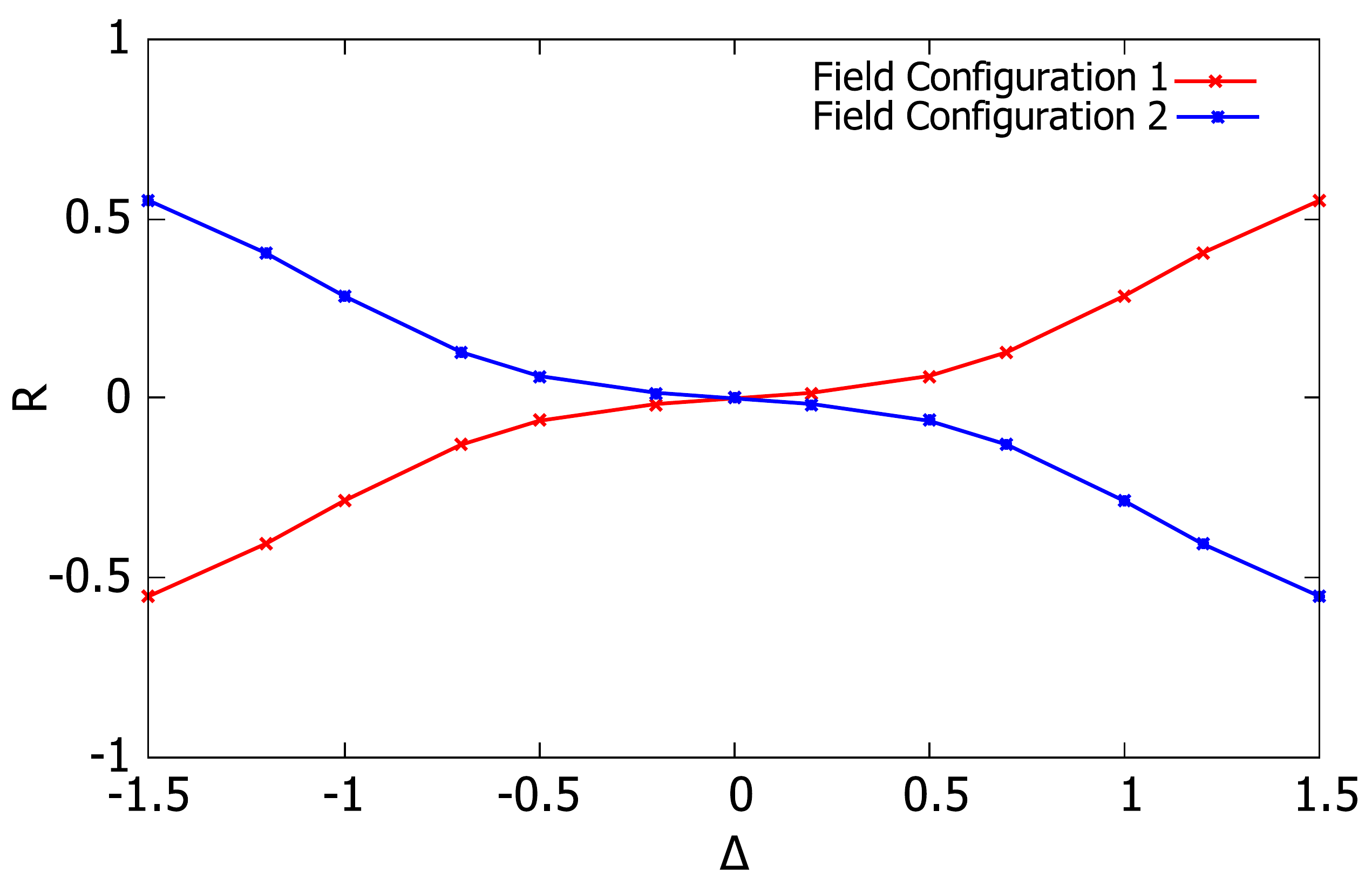}
    \caption{Rectification coefficient $R$ vs $\Delta$ for the symmetrical lattice with nine sites (Fig. \ref{g4}), under two configurations of non-homogeneous magnetic fields.}
    \label{g4plot}
\end{figure}

For the third lattice (Fig. \ref{g3}) with a symmetric geometry, under a magnetic field varying from $h=1$ to $h=5$ and $\gamma_{j \in R} = \gamma_{j \in L}=1.0$, the results are in Fig. \ref{g3plot}. And finally,  the  results for the fourth lattice (Fig. \ref{g4}), under a  field varying from $h=1$ to $h=3$ and $\gamma_{j \in R} = \gamma_{j \in L}=1.0$, are in Fig. \ref{g4plot}.
The symmetric geometries shows symmetric values of rectification around $\Delta =0$ where $R=0$ and have a similar behavior as the one-dimensional cases, where the XX model does not present rectification even under a non-homogeneous magnetic field. It is important to emphasise that the inversion of the field configurations just changes the sign of the rectification coefficient, as expected for the symmetric cases.

For all geometries we expect that the rectification reaches a maximum point and then starts to decreases as $|\Delta|$ increases.

Additionally, it is interesting to point out that these kinds of geometries may allow interference effects to play a role. For instance, if we consider the XX model in the geometry in Fig.~\ref{g1}, without the reservoirs, an excitation initially put in the right column, in a suitable superposition, may never reach the leftmost site.  The reservoirs we have considered may obscure these effects, since they pump particles in the right column incoherently.

To verify interference effects, we consider two types of collective reservoirs on the right side for the triangular lattice with $10$ sites (Fig. \ref{g1}). The first reservoir is described by: 
\begin{equation}
    L_{R, \pm} = \sqrt{1 \pm f_{R}} \;(\sigma_{7}^{\pm} + \sigma_{8}^{\pm} +\sigma_{9}^{\pm}+\sigma_{10}^{\pm}).
\end{equation}
And for the second one alternate phases are included:

\begin{equation}
    L_{R, \pm} = \sqrt{1 \pm f_{R}}\; (\sigma_{7}^{\pm} +  i\; \sigma_{8}^{\pm} +\sigma_{9}^{\pm}+ i \;\sigma_{10}^{\pm}).
\end{equation}

However, at least for the system sizes we were able to consider, both collective reservoirs were not able to show significant differences in the rectification behavior, compared to the cases with separated reservoirs.

Furthermore, to certify that the geometry alone does not induces the rectification, two extra cases with  six sites and homogeneous magnetic field ($h=1.0$) are considered: In the first situation we fix one reservoir on the left side and two on the right side; in the second case we fix one reservoir on the left and three on the right. Then, all possible geometries are tested, including cases where sites and reservoirs are excluded. None of them exhibited rectification, indicating that a non-homogeneous magnetic field is necessary to the occurrence of magnetic currents rectification.


\section{Conclusions} \label{sec5}

We have explored numerically the rectification $R$ in two symmetric and two asymmetric geometries for two-dimensional XX and XXZ spin lattices, under one homogeneous and  two non-homogeneous magnetic field configurations. In addition, we also considered two collective reservoirs for the triangular lattice with ten sites and all possible geometries with six sites under a homogeneous magnetic field.   

The results have shown that the non-homogeneous magnetic field is a required element, but besides that, the interesting occurrence of rectification in the XX model for asymmetric geometries distinguish its behavior from the symmetric cases considered, that are similar to one-dimensional chains.  

Finally, we conclude that these outcomes described for archetypal models of open quantum system suggest the geometry can interfere significantly on spin transport and, particularly, its asymmetry reveal new possibilities for inducing rectification in quantum spin lattices, even in the  XX model. Importantly, the occurrence of spin rectification in such simple models indicates that the phenomenon is of general occurrence in spin systems.

We hope our results stimulate more research on the theme of geometry-induced rectification,
in particular, it will be interesting to investigate these effects in the energy current as well.


\begin{acknowledgments}
This study was financed in part by the Coordenação de Aperfeiçoamento de Pessoal de Nível Superior - Brasil (CAPES) - Finance Code 001.

E.P. was partially suported by CNPq (Brazil).
\end{acknowledgments}

\nocite{*}

\bibliography{apssamp}

\end{document}